\documentstyle[preprint,revtex]{aps}
\begin{document}
\def\arcsinh{\mathop{\rm arcsinh}\nolimits}
\rightline{IUHET-241}
\rightline{IPS Research Report No. 92-29}
\rightline{UALG-PHYS-12}
\begin{title}
A realistic heat bath:

theory and application to kink-antikink dynamics
\end{title}
\author{A.~Krasnitz* and Robertus Potting**}
\begin{instit}
Physics Department

Indiana University

Bloomington, Indiana 47405

and

*Interdisciplinary Project Center for Supercomputing

Eidgen\"ossische Technische Hochschule-Zentrum

CH-8092 Zurich, Switzerland

**Universidade do Algarve

Unidade de Ci\^encias Exactas e Humanas

Campus de Gambelas, 8000 Faro, Portugal
\end{instit}
\begin{abstract}
We propose a new method of studying
a real-time canonical evolution of field-theoretic systems
with boundary coupling to a realistic heat bath.
In the free-field case the method is equivalent to an infinite extension
of the system beyond the boundary, while in the interacting case
the extension of the system is done in linear approximation.
We use this technique to study kink-antikink dynamics
in $\varphi^4$ field theory in 1+1 dimensions.
\end{abstract}
\newpage
\def\half{{\textstyle 1\over2}}
\section{Introduction}
There is a growing interest in
transitions over energy barriers in field theories at finite temperature.
The motivation comes from high-energy physics and cosmology
({\it e.g.,} baryon-number violating sphaleron transitions
in electroweak theory\cite{ML}-\cite{GR}), as
well as from condensed-matter physics ({\it e.g.,} current-reducing
fluctuations
in one-dimensional superconductors\cite{NG}). Such thermally induced
transitions
usually involve collective excitations like kink-antikink pairs or
sphalerons, and therefore occur in a nonlinear setting. The nonlinearity,
in turn, often renders analytic techniques useless, forcing one to resort
to numerical (lattice) methods. On
the other hand, many questions of interest can be answered by
considering the classical regime in which
the temperature is much larger than the
energies of the field quanta at the scale of a relevant collective
excitation\cite{GR}.

A fundamental quantity of interest is the corresponding transition rate.
Unlike static observables, it cannot be determined from a given canonical
ensemble of field configurations. Instead, one must follow a real-time
evolution of the field-theoretic system in the heat bath. Obviously,
the rate would depend on the properties of the heat bath and its coupling
to the system. This was indeed confirmed by recent numerical studies\cite{BdF2}
\cite{SB}, in which the heat bath was implemented through Langevin equation.
In most
naturally occurring situations a system interacts with its environment
through the boundaries, and the environment is an infinite extension
of the system itself. This consideration dictates our choice of a heat
bath in this work.
Ideally, the heat bath we model
should possess two main properties: ({\it a})
waves traveling through the boundary out of the system should be completely
absorbed, as if there were no boundary and the system had an infinite
extension;
and ({\it b}) the waves traveling into the system should be thermally
distributed. Our construction of a heat bath ensures these two properties
in the free-field case, while
for an interacting field they
hold in the linear approximation.
The similarity between our linearized heat bath and the
true extension of the interacting system becomes better with
decreasing temperature. Note that the standard Langevin dynamics, imposed
either
in the bulk of the system or at the boundary\cite{ZT}, will not lead to
the properties {\it a} and {\it b}.

While the idea of mimicking a natural heat bath is very generally applicable,
we restrict ourselves in this article to the case of
a scalar field in one spatial dimension. In Section II we derive the
boundary force exerted by the heat bath on this one-dimensional system.
Next, we show in Section III how our construction of
the realistic heat bath can be
implemented numerically. As an example, in Section IV we apply our technique
to study kink-antikink pair nucleation in $\varphi^4$ theory, a subject that
has been
extensively investigated by other methods\cite{SB}-\cite{GR}.
We summarize and discuss our results
in Section V.

\section{The method}
Consider a lattice
system consisting of a (possibly self-interacting) scalar field
$\varphi_n$, with a minimum of its potential $V(\varphi)$
at $\varphi=v$ and mass $m$
corresponding to that minimum.
For definiteness, let it reside on the $n\le0$ sites of a one-dimensional
chain,
and be coupled to a
heat bath at $n=0$. Let the heat bath be a free
massive scalar field system on the positive $n$ sites with equation of motion
(the lattice spacing is $a$)
\begin{equation}
\ddot\Phi_n-{\Phi_{n+1}+\Phi_{n-1}-2\Phi_n \over a^2}+m^2\Phi_n
=0
\label{eom}
\end{equation}
for $n\ge1$ and boundary condition
\begin{equation}
\Phi_0(t)=\varphi_0(t)-v=f(t).
\label{boundary}
\end{equation}
Given the solution $\Phi_n(t)$, there will be a reaction force exerted by
the heat bath on the system:
\begin{equation}
F(t)=(\Phi_1-\Phi_0)/a^2.
\label{reac}
\end{equation}
The $\Phi_0$ contribution to $F(t)$ is a harmonic force. The $\Phi_1$
contribution, called $f_{\rm mem}$ (the memory force) in the following, is, as
we shall see immediately, a response of a more general nature.
As the heat bath is a free field system, this response
should be linear. Furthermore, it should causally depend on $f(t)$.
In other words, there exists a response function $\chi(t)$ $(t>0)$ such that
\begin{equation}
f_{\rm mem}(t)=\int_{-\infty}^t f(t')\chi(t-t')dt'.
\label{response}
\end{equation}
In order to determine $\chi(t)$ let us take $f(t)=e^{st}$. It follows
from Eq.~\ref{eom} that $\Phi_n(t)=\exp(st-k_s an)$, with
\begin{equation}
k_s={2\over a}\arcsinh\left({a\over2}\sqrt{s^2+m^2}\right).
\end{equation}
Using Eq.~\ref{reac} we then find that $\chi(t)$ should satisfy
\begin{equation}
{1\over a^2}e^{st-k_sa}
=\int_{-\infty}^t e^{st'}\chi(t-t')dt'=e^{st}({\cal L}\chi)(s),
\label{linear}
\end{equation}
where ${\cal L}$ denotes the Laplace transform. It follows that
\begin{eqnarray}
\chi(t)&=&a^{-2}{\cal L}^{-1}(e^{-ak_s})\nonumber\\
&=&{1\over a^2}{\cal L}^{-1}\left((\sqrt{1+y}-\sqrt{y})^2\right)
\qquad\hbox{with }y=a^2(s^2+m^2)/4.
\label{LaplacePsi}
\end{eqnarray}
The integral in Eq.~\ref{response} is most easily computed
in Fourier space.
We define $\tilde f_t(\omega)=\int_{-\infty}^t e^{i\omega(t'-t)}f(t')dt'$ and
$\tilde\chi(\omega)=\int_{-\infty}^\infty e^{i\omega t}\chi(t)dt$,
where we are free to take any extension of $\chi(t)$ for $t<0$.
Then it follows from the convolution theorem that
\begin{equation}
f_{\rm mem}(t)={1\over2\pi}\int \tilde\chi(\omega)\tilde
f_t(\omega) d\omega.
\label{fmem}
\end{equation}
For numerical purposes it is best to extend $\chi(t)$ to $t<0$
such that $\tilde\chi(\omega)$ vanishes outside a finite interval. To this end
we choose $\chi(t)=-\chi(-t)$,
so $\tilde\chi(\omega)$ becomes purely imaginary (essentially the sine
transform).
The latter can be obtained immediately from the Laplace transform
(Eq.~\ref{LaplacePsi}) by the $\pi/2$ rotation of $s$ in the complex plane;
taking the imaginary part then gives
\begin{equation}
\tilde\chi(\omega)=
i{\rm sign}(\omega)\sqrt{(\omega^2-m^2)(1+a^2(m^2-\omega^2)/4)}
\label{tildechi}
\end{equation}
for frequencies $m<|\omega|<\sqrt{m^2+4/a^2}$
corresponding to
propagating modes and $\tilde\chi(\omega)=0$ outside this range.

We expect the heat bath response to be finite for any bounded $f(t)$
(Eq.~\ref{response}). If so, $\chi(t)$ must decay sufficiently rapidly in the
distant future. This indeed is the case: using Eq.~\ref{tildechi} and
performing
the inverse Fourier transform we find in the saddle-point approximation
$\chi(t)\propto t^{-3/2}\sin(mt+\pi/4)$ as $t\rightarrow\infty$.

The dissipation by the heat bath can equivalently be described by a
linear response to the momentum $\pi_0$ at the boundary, rather than the
field. Explicitly, it follows from Eq.~\ref{response} that
\begin{equation}
f_{\rm mem}(t)=-\int_{-\infty}^t \psi(t-t')\pi_0(t')dt',
\end{equation}
where $\psi(t)=\int_0^t \chi(t')dt'$ or, equivalently,
$({\cal L}\psi)(s)=s^{-1}({\cal L}\chi)(s)$, so that
\begin{equation}
\tilde\psi(\omega)=i\omega^{-1}\tilde\chi(\omega).
\label{PsiChi}
\end{equation}

We now turn to the second contribution to the boundary force, the
random force $f_{\rm ran}$. It is a Gaussian random variable whose
properties are defined by its time autocorrelation $C(\tau)=\langle
f_{\rm ran}(t)f_{\rm ran}(t+\tau)\rangle$. These are most easily determined
through the fluctuation-dissipation theorem, stating
\begin{equation}
C(\tau)=\theta\psi(|\tau|)
\label{Ct}
\end{equation}
($\theta$ denotes the temperature).
Numerical implementation of $f_{\rm ran}$
amounts to generating Gaussian noise with this
autocorrelation function.

One can check explicitly that the time correlation function
of the total boundary force $F+f_{\rm ran}$ is equal to the
average of
$((\Phi_n(0)-\Phi_{n-1}(0))((\Phi_n(t)-\Phi_{n-1}(t))/a^4$
for a canonical ensemble, calculated for an infinitely extended free
field. This is to be expected, as $(\Phi_n-\Phi_{n-1})/a^2$ represents
the force between neighboring sites.

We conclude this section by discussing the continuum limit of the heat bath
response. Obviously, $F(t)$ diverges as $a\rightarrow 0$. Instead, $aF(t)$
is a well-behaved quantity whose limit is simply $\partial_x\Phi(x=0,t)$,
where $x$ is the continuum spatial coordinate. In
other words, if we prescribe the boundary field motion, the heat bath response
will determine the boundary field spatial derivative in such a way that the
waves traveling out of the system will not be reflected at the boundary.
The corresponding response function is found following steps analogous
to Eq.~\ref{response}--Eq.~\ref{LaplacePsi}. We write down
the continuum equations of motion
\begin{equation}
\ddot\Phi-\partial_x^2\Phi+m^2\Phi=0\qquad(x>0)
\label{conteom}
\end{equation}
together with the boundary condition $\Phi(0,t)=f(t)$ and require
\begin{equation}
\partial_x \Phi(0,t)=\int_{-\infty}^t f(t')\chi_c(t-t')dt'.
\label{contresponse}
\end{equation}
Taking $f(t)=\exp(st)$ then yields $\chi_c(t)={\cal L}^{-1}(\sqrt{s^2+m^2})$,
or $\chi_c(t)=\delta'(t-0_+)+mt^{-1}J_1(mt)$. It is an easy exercise to verify
that the same result is obtained by first computing the response function for
the lattice boundary field derivative and then taking the limit
$a\rightarrow 0$.

\section{Numerical implementation}
Our numerical implementation of the boundary
force is as follows. The memory force
$f_{\rm mem}(t)$ is computed using Eq.~\ref{fmem}, that is, in Fourier
space. The function $\tilde\chi(\omega)$ is given by
Eq.~\ref{tildechi}. In practice, the integral Eq.~\ref{fmem} is replaced
by a finite sum over discrete values of $\omega$, separated by an
increment $\Delta\omega$. As a result, the response function $\chi(t)$ becomes
periodic in $t$ with a period $2\pi/\Delta\omega$.
Beyond $t=2\pi/\Delta\omega$ our approximation of $\chi(t)$ is incorrect,
and the lower limit of integration in Eq.~\ref{response} should be cut off.
We achieve this by changing the definition of
$\tilde f_t(\omega)$ to
\begin{equation}
\tilde f_t(\omega)=\int_{t-T}^t e^{i\omega(t'-t)}f(t')dt',
\label{ftomega}
\end{equation}
where $T<2\pi/\Delta\omega$. At the same time, $T$ should be large
enough so the discarded part of the integral Eq.~\ref{fmem} is
negligible.
With the new definition, $\tilde
f_t(\omega)$ satisfies the equation of motion
\begin{equation}
\dot{\tilde f_t}(\omega)=-i\omega\tilde f_t(\omega)+f(t)-e^{-i\omega T}f(t-T).
\label{updateft}
\end{equation}

The random force $f_{\rm ran}$ is computed by convolving white noise of
unit power spectrum with
a function $R(t)$ whose Fourier image is given by
$\sqrt{\tilde C(\omega)}$, where $\tilde C$ is the Fourier transform
of $C(t)$ given by Eq.~\ref{Ct} \cite{Rt}.
This can be done very efficiently using FFT algorithms.

With both ingredients of the boundary force in place, we can now write
down and integrate the equations of motion for the one-dimensional
system immersed in the heat bath. These equations have the standard form
\begin{equation}
\ddot\varphi_n - {\varphi_{n+1}+\varphi_{n-1}-2\varphi_n \over a^2} -
V'(\varphi_n)=0
\label{eq:eqmo_bulk}
\end{equation}
in the bulk of the system, but should be modified at the boundaries. For
example, at the left boundary we have
\begin{equation}
\ddot\varphi_0 - {{\varphi_1-2\varphi_0}\over a^2} - V'(\varphi_0)=
f_{\rm mem,left}+f_{\rm ran,left},
\label{eq:eqmo_bound}
\end{equation}
and the right-boundary analog is obvious. The system of
equations of motion for the field is supplemented by
Eq.~\ref{updateft} governing the evolution of the memory forces at
both boundaries.

We integrate equations \ref{eq:eqmo_bulk}, \ref{eq:eqmo_bound},
and \ref{updateft} using the
second-order Runge-Kutta algorithm. While this way of updating
$\tilde f_t(\omega)$ is efficient compared to computation of the full
integral Eq.~\ref{ftomega}, it is not accurate enough to maintain correct
phases of $\tilde f_t(\omega)$ for times much longer than $2\pi/\Delta\omega$.
To ensure stability, we therefore adjust the values of $\tilde f_t(\omega)$ by
computing Eq.~\ref{ftomega} every $T$ time units.

We have tested the action of $f_{\rm mem}$
by evolving an initially hot system with
$f_{\rm ran}$ omitted. This corresponds to cooling in the zero-temperature heat
bath. The resulting
evolution is very similar to that of a benchmark run where we
substitute for the heat bath a real cold free-field system with large
volume. This similarity holds separately for every momentum mode.
The method allows cooling the system to at least $10^{-5}$ times its original
temperature. Cooling curves for a system with a real and with a simulated heat
bath are shown in Figure \ref{coolcomp}.

We then turned $f_{\rm ran}$ on and tested it
by heating a cold field configuration to a
prescribed temperature. Again it compared well, mode by mode, with a
benchmark run, in which we coupled a cold system to a large real heat bath
whose
temperature the system eventually reached
(let us stress that using the real heat bath is much more costly in terms of
CPU time). Both heating curves are shown in Figure \ref{heatcomp}.

Finally, we have shown numerically the self-consistency of our method by
comparing the motion of the endpoint field to that in the middle of the system
\cite{PhdF}. As Figure \ref{endvsmid}
shows, the autocorrelation
curves of the two fields are very close to each other, meaning that the
simulated free-field heat bath closely approximates the real one.

\section{Kink-antikink dynamics in 1+1 dimensions}
We have applied our method to investigate
kink-antikink dynamics of $\varphi^4$ theory whose Lagrangian is
\begin{equation}
L=\int dx \left(\half\left(\dot\varphi^2-(\partial_x\varphi)^2\right) -
{1\over 4}(\varphi^2-1)^2\right)
\end{equation}
in suitably chosen units\cite{GR}.
Kinks and antikinks are finite-energy solutions of
the equations of motion interpolating between the vacuum values of the field
$\varphi=\pm1$. Explicit functional form of the static kink is
$\varphi_\pm(x)=\tanh(\pm {x\over\sqrt{2}})$. The spatial extension of a kink
is approximately $\sqrt{2}$, and the kink mass is $M=\sqrt{8/9}$.

Following \cite{SB}, we chose a system of
$N=400$ sites with lattice spacing $a=0.5$.
The simulations were performed at seven values of
inverse temperature $\beta$ between 3.0 and 6.0.
Our Runge-Kutta time step was 0.005. For each
value of the temperature we started from an ordered system and warmed it
up for 2500 time units followed by $2\times 10^5$ time units over which
we measured the number of kinks $n$. For the latter we used the same definition
as in \cite{SB}: the number of zeros in a field
configuration smoothened over the physical distance of $\Delta L=5$.

The average kink number $\langle n \rangle$
at a given temperature should not depend on the
properties of a heat bath. Our measurements of this observable, shown in Figure
\ref{kinkno_vs_beta},
are indeed close to those of \cite{SB}. Note, however, that our measurement
errors are much larger, especially for low temperatures, even though our data
sample is as big as in \cite{SB}. We estimated the
errors using a jackknife method with the block size varying over a very long
range; we are therefore confident that the autocorrelation of our data is
properly taken into account. Moreover, we studied a microcanonical evolution
of our system at the energy roughly corresponding to $\beta=4.5$, with the
error estimate similar to that of the corresponding canonical case.

The temperature dependence of $\langle n\rangle$
may be interpreted in terms
of the effective kink mass. Namely, one expects\cite{HMS}
\begin{equation}
\langle n\rangle\propto\sqrt{\beta}\exp(-\beta M_{\rm eff}).\label{eq:meff}
\end{equation}
It was found in earlier work that $M_{\rm eff}<M$. It was also indicated that
$M_{\rm eff}$ is temperature dependent\cite{SB},\cite{BdF}.
Both features find further evidence
in our study. If we try to fit all our measurements of $\langle n \rangle$
to Eq.~\ref{eq:meff} at once, an unacceptably low goodness-of-fit results. The
situation improves dramatically if we exclude the highest-temperature point
from the fit. We then find $M_{\rm eff}=0.695\pm 0.0095$,
or $M_{\rm eff}=(0.737\pm 0.010)M$, in good agreement with
\cite{SB}. Alternatively, we can use pairs of consecutive values of
$\langle n \rangle$ to extract $M_{\rm eff}$. The result, presented in Figure
\ref{meff2p},
shows the tendency of $M_{\rm eff}$ to decrease at higher temperatures, in
agreement with findings of \cite{SB}, \cite{BdF}.

Another interesting quantity we extract from the kink-antikink number time
history is a kink lifetime, {\it i.e.}~the autocorrelation time $\tau$
of $n$. The latter is usually obtained by fitting the $n$ autocorrelation
function $\langle(n(t)-\langle n\rangle)(n(0)-\langle n\rangle)\rangle$
to a single exponential of the form $\exp(-t/\tau)$. If the time history
exhibits more than one time scale (as is the case for $n$), $\tau$ can only be
given an average, or effective meaning. The existence of such multiple scales
also makes a single-exponential fit to the autocorrelation function extremely
difficult. A
multiexponential fit to noisy data is not a practical possibility.
We use an alternative way of determining $\tau$, closely related to
the integral definition of the autocorrelation time. Namely, if $\Delta t$
is a time interval between two consecutive
measurements of $n$, we expect
\begin{eqnarray}
w(N)&\equiv&
\langle(\sum_{i=0}^{N-1}n(i\Delta t)-N\langle
n\rangle)^2\rangle\label{deftau}\\
&\approx&\langle(n-\langle n\rangle)^2\rangle
\left[{N\over{\tanh\left({{\Delta t}\over{2\tau}}\right)}}-{1\over{2\sinh^2
\left({{\Delta t}\over{2\tau}}\right)}}\right].\nonumber
\end{eqnarray}
Obviously, for large $N$ $w(N)$ approaches a random-walk behavior.
For a given $N$ we determine $w(N)$ from our data set and solve
Eq.~\ref{deftau}
for $\tau$. For $t\equiv N\Delta t>>\tau$
the result is approximately independent of
$N$, and we take it as an estimate of the kink lifetime. A typical dependence
of $\tau$ on $t$ is shown in Figure \ref{tau_vs_lag}.
Note that large values of $t$ for which
the plateau is reached indicate the existence of multiple time scales in the
kink-antikink number fluctuations.

Unlike the
exponential fit, this method also allows a well-defined error-estimating
procedure for $\tau$. In particular, we apply the jackknife technique. Note
that the lag values $t$ used to determine $\tau$ are of the order of 5000.
This is to be compared to our $n$ time history length of $2\times 10^5$.
We therefore only have a small effective number of independent measurements
of $\tau$, and our error estimate cannot be very accurate. Conservatively
we can expect the errors of $\tau$ to be correct within a factor of 2.
This might explain their
inhomogeneous dependence on the temperature.

Kink-antikink pair nucleation can be viewed as a multidimensional analog of
a particle escape over a barrier\cite{Dine},\cite{HMS}, with the kink lifetime
related to to the effective barrier height $B$\cite{SB}:
\begin{equation}
\tau\propto\exp\left(\beta U\right),\label{taupred}
\end{equation}
where $U=B-M_{\rm eff}$.
{}From our data (Figure \ref{tau_vs_beta}) we find
$U=1.01\pm0.06=(1.07\pm0.06)M$,
slightly
higher than $U=(0.85\pm0.15)M$ of \cite{SB}, obtained by solving a
low-viscosity
Langevin equation. Both that value and ours
are inconsistent with the naively expected $B=2M$.
More work is required to explain this discrepancy\cite{DL}.
More importantly, however,
our results show no exponential suppression of the kink-antikink pairs
nucleation rate with growing temperature, in agreement with analytical
predictions and earlier numerical work\cite{Dine}-\cite{GR}.
\section{Conclusions and Outlook}
In this article we presented a method of modeling naturally occurring heat
baths. While our presentation concentrated on a scalar field in one spatial
dimension, the principles underlying our construction of a heat bath do not
depend on the dimensionality or the field content of a theory in question.
In any case, one can determine the memory force exerted by a heat bath by
studying the linear response of the latter to the field motion at the system
boundary. The corresponding random component of the force may then be found
using fluctuation-dissipation theorem. The only new feature to appear in
dimensions higher than one is related to the connectedness of the boundary:
the field motion at different points of the boundary will be correlated in a
way
consistent with causality. This is, however, a technical difficulty,
not a conceptual one. The work on extending our method to other systems
is currently in progress.

We have verified numerically that our simulated heat bath thermalizes correctly
both linear and nonlinear systems. As an application, we considered the
dynamics of kink-antikink pairs in $\varphi^4$ theory. Our measurements of the
kink density agree well with those obtained by solving Langevin equation, as
one would expect for an equilibrium quantity independent of a heat bath
implementation. The kink lifetimes we measure are close to those following
from the low-viscosity Langevin dynamics. This is again to be expected for
sufficiently large systems: as the system size grows, the influence of the
boundary heat bath on the dynamics decreases. The Langevin analog would then be
decreasing viscosity.

To conclude, we emphasize again
an important advantage of our simulated heat bath over the Langevin
method (including its zero-viscosity microcanonical limit): our heat bath is
not arbitrarily chosen and involves no free parameters like viscosity.
Rather, it is, to the best of our knowledge, the closest known approximation
to a natural situation, in which open systems are immersed in a similar
environment.
The only physical (not numerical)
approximation we make is linearization: our
heat bath exchanges linear excitations (plane waves, or mesons) with the
system,
but not nonlinear ones (kinks, or baryons). Since we are only interested in
pair creation and annihilation processes, the exchange of topological charge
with the environment should have little impact on our results.
It follows that we can approximately (the only approximation
being linearization of the heat bath) identify our simulation time with
the real physical one. The transition rates we
measure have therefore direct physical meaning.
\nonum
\section{Acknowledgments}

We are indebted to Ph.~de Forcrand, to S.~Gottlieb, and to A.~Kovner
for many interesting
discussions and helpful suggestions. This work was supported by the US
Department of Energy and by the Swiss Nationalfond.

\newpage
\figure{Cooling curves for a free-field system ($m=0.5, a=1, L=100$)
immersed in a simulated (squares) and real (solid line) zero-temperature
heat baths.\label{coolcomp}}
\figure{Heating curves for a free-field system ($m=0.5, a=1, L=100$)
immersed in a simulated (squares) and real (solid line) heat baths.
The equilibrium is reached at $\theta=1$.\label{heatcomp}}
\figure{Autocorrelation functions of the endpoint (solid line) and midpoint
(squares) fields for a free-field system
($m=0.5, a=1, L=100$) at $\theta=1$.\label{endvsmid}}
\figure{Temperature dependence of the kink number (logarithmic scale). The
solid line is a fit for $M_{\rm eff}=0.737 M$.\label{kinkno_vs_beta}}
\figure{Temperature dependence of the kink effective mass obtained by fitting
pairs of consecutive points from Figure \ref{kinkno_vs_beta} to
Eq.~\ref{eq:meff}.
The inverse temperature $\beta$ is that of a higher-temperature point in each
pair.\label{meff2p}}
\figure{The kink number autocorrelation time $\tau$ at $\beta=4$ determined
from
Eq.~\ref{deftau}, plotted as a function of the lag $N\Delta t$.
\label{tau_vs_lag}}
\figure{Temperature dependence of the kink lifetime (logarithmic scale). The
solid line is a fit for $U=1.06M$.\label{tau_vs_beta}}
\newpage
\nonum\section{Figure 1}
\includegraphics{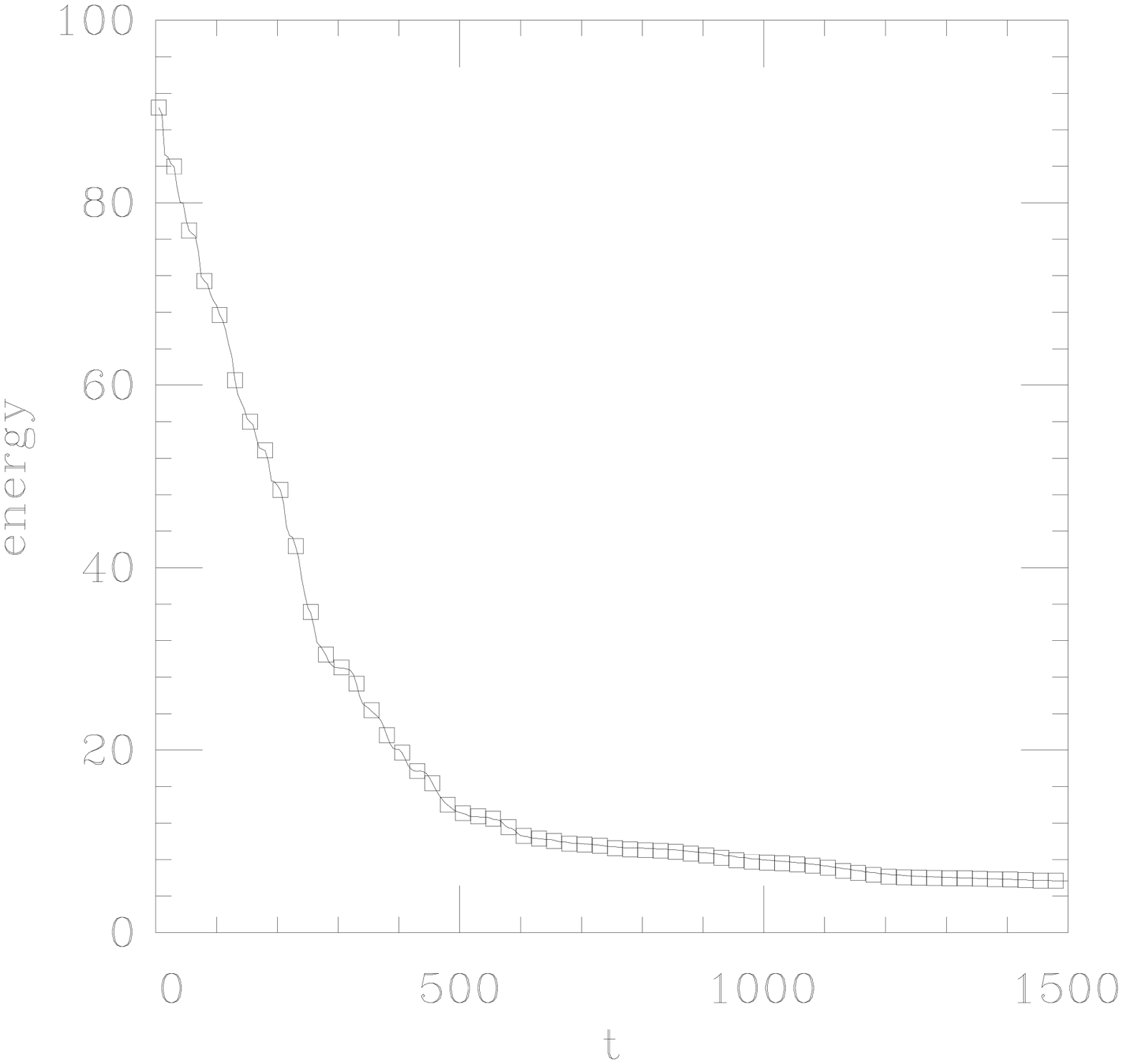}
\newpage
\nonum\section{Figure 2}
\includegraphics{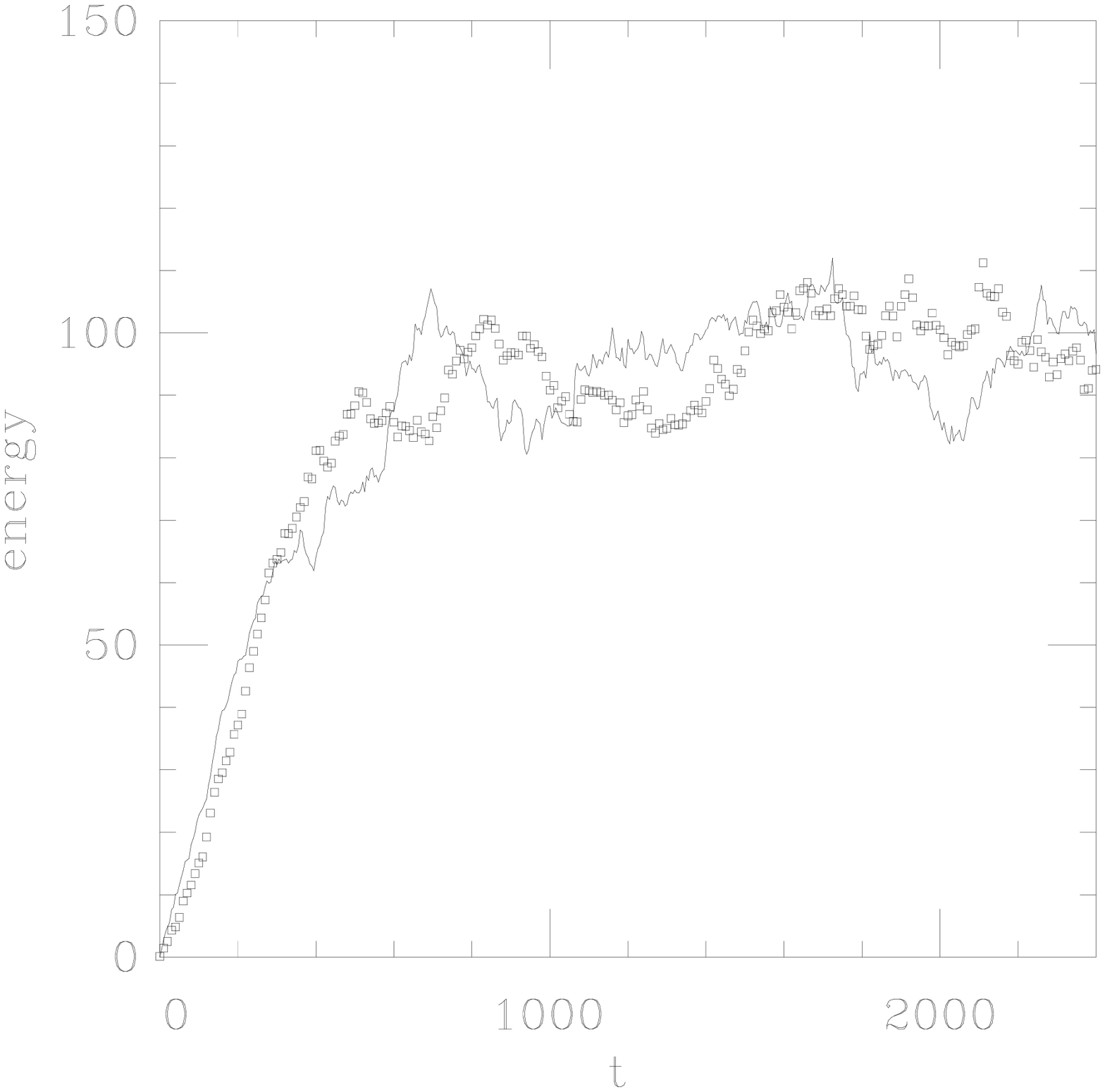}
\newpage
\nonum\section{Figure 3}
\includegraphics{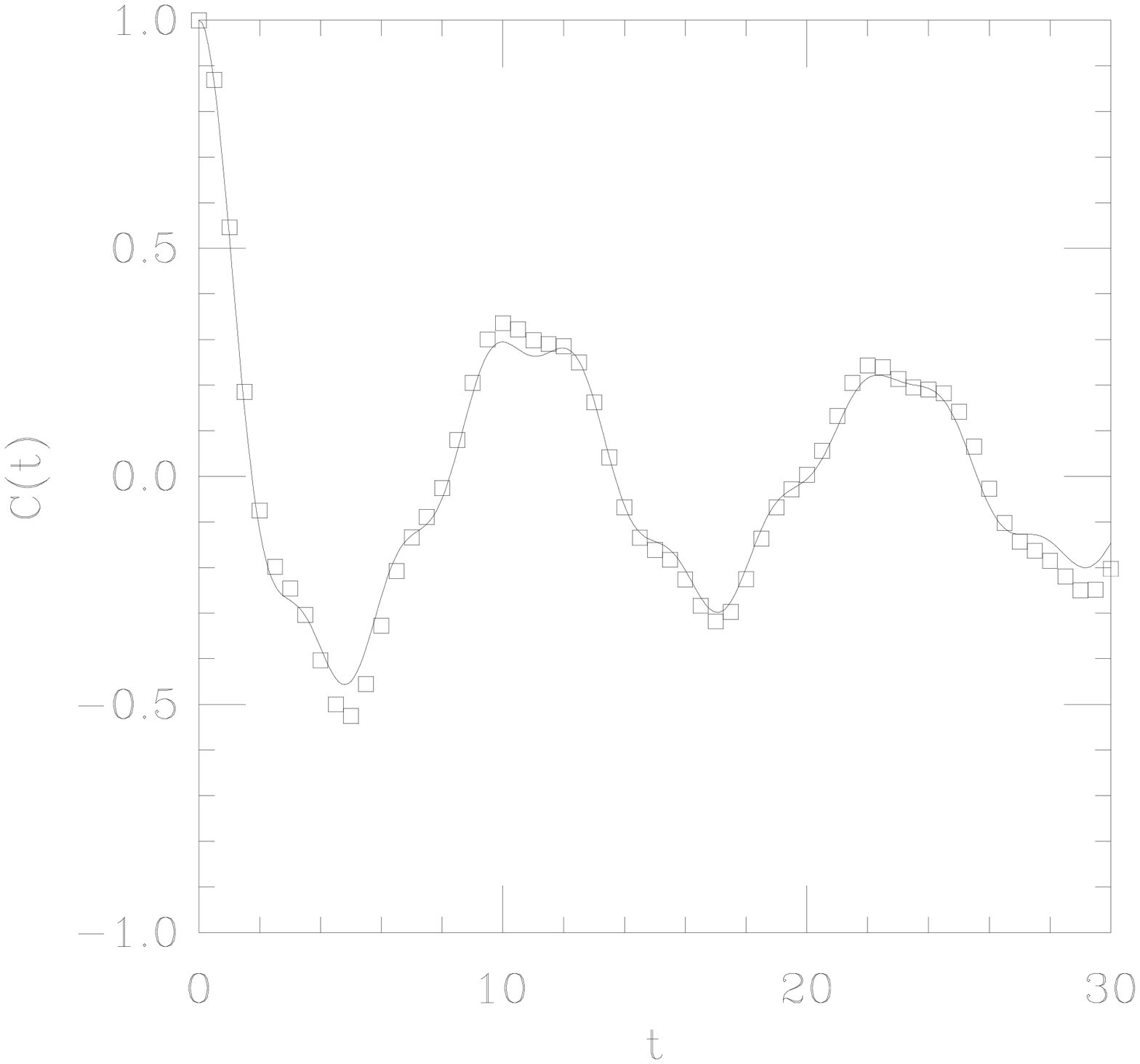}
\newpage
\nonum\section{Figure 4}
\includegraphics{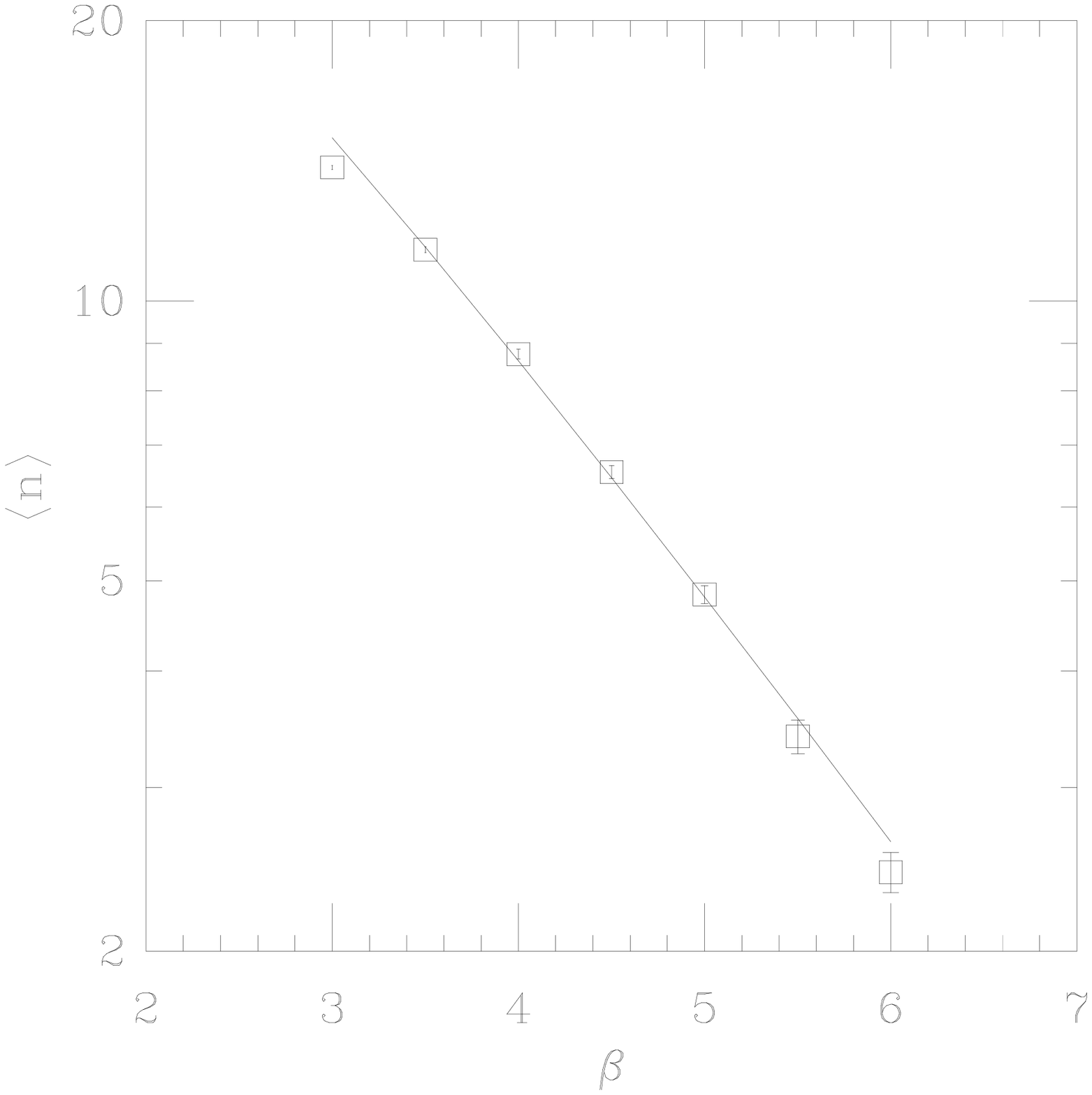}
\newpage
\nonum\section{Figure 5}
\includegraphics{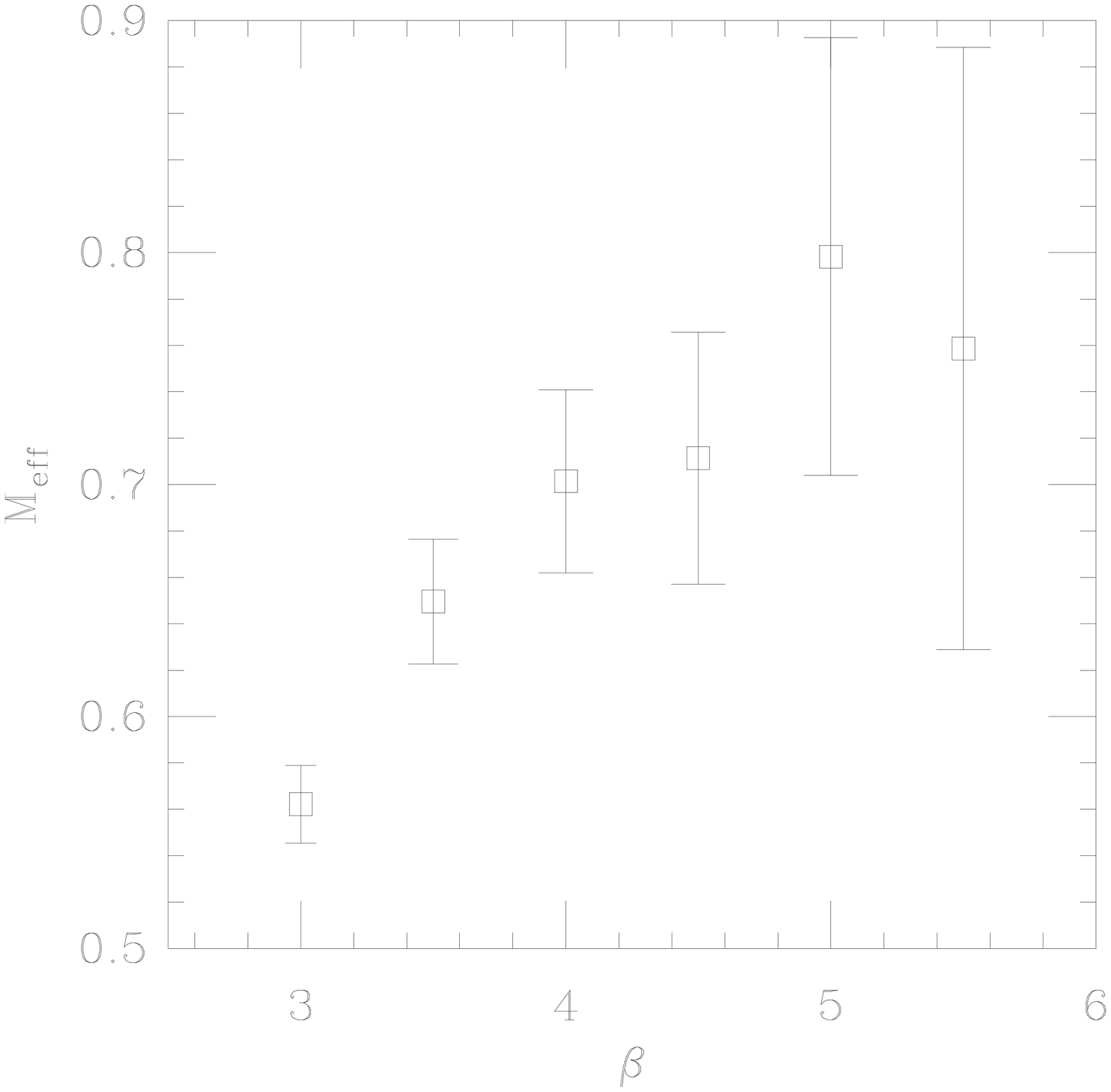}
\newpage
\nonum\section{Figure 6}
\includegraphics{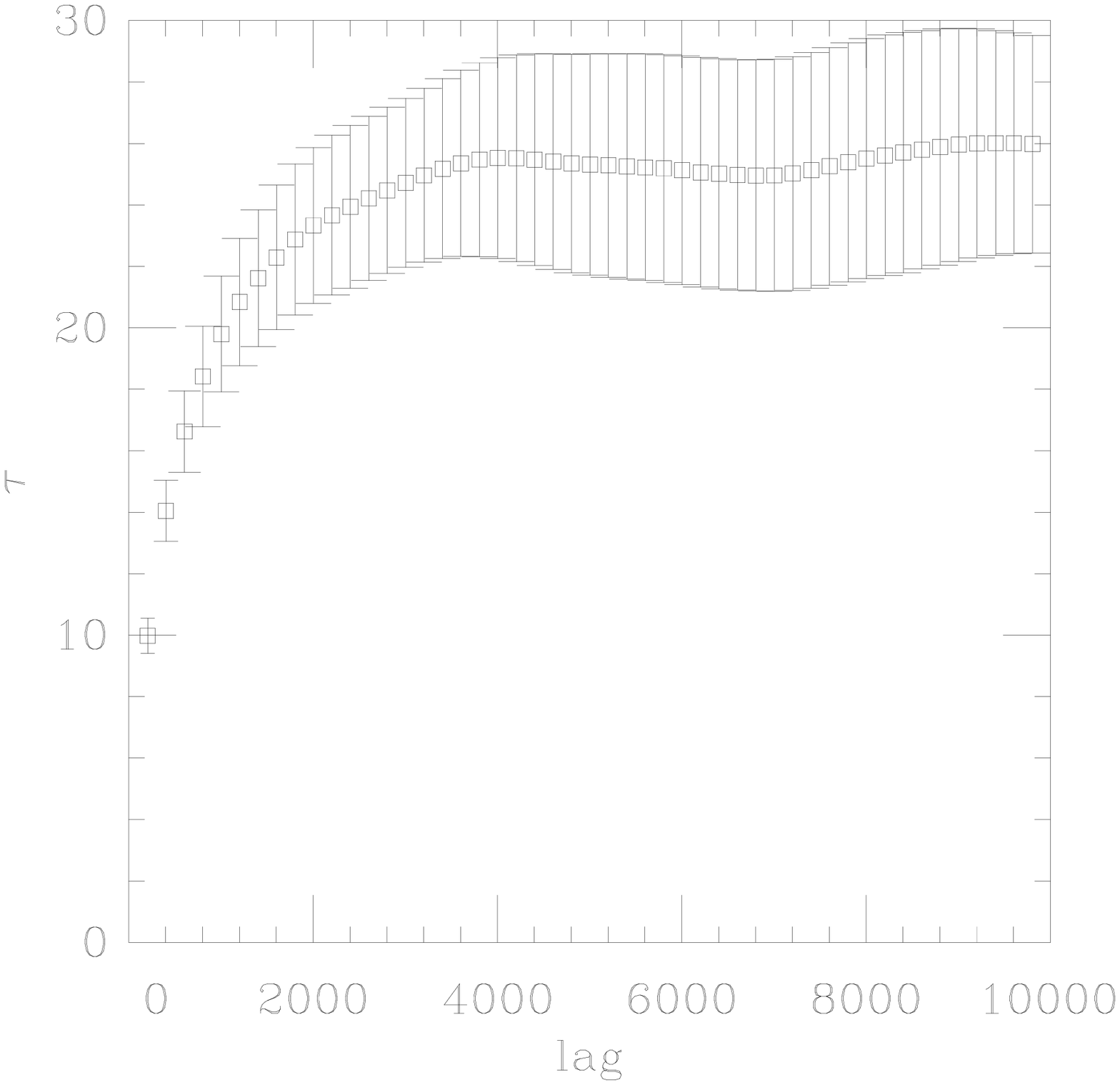}
\newpage
\nonum\section{Figure 7}
\includegraphics{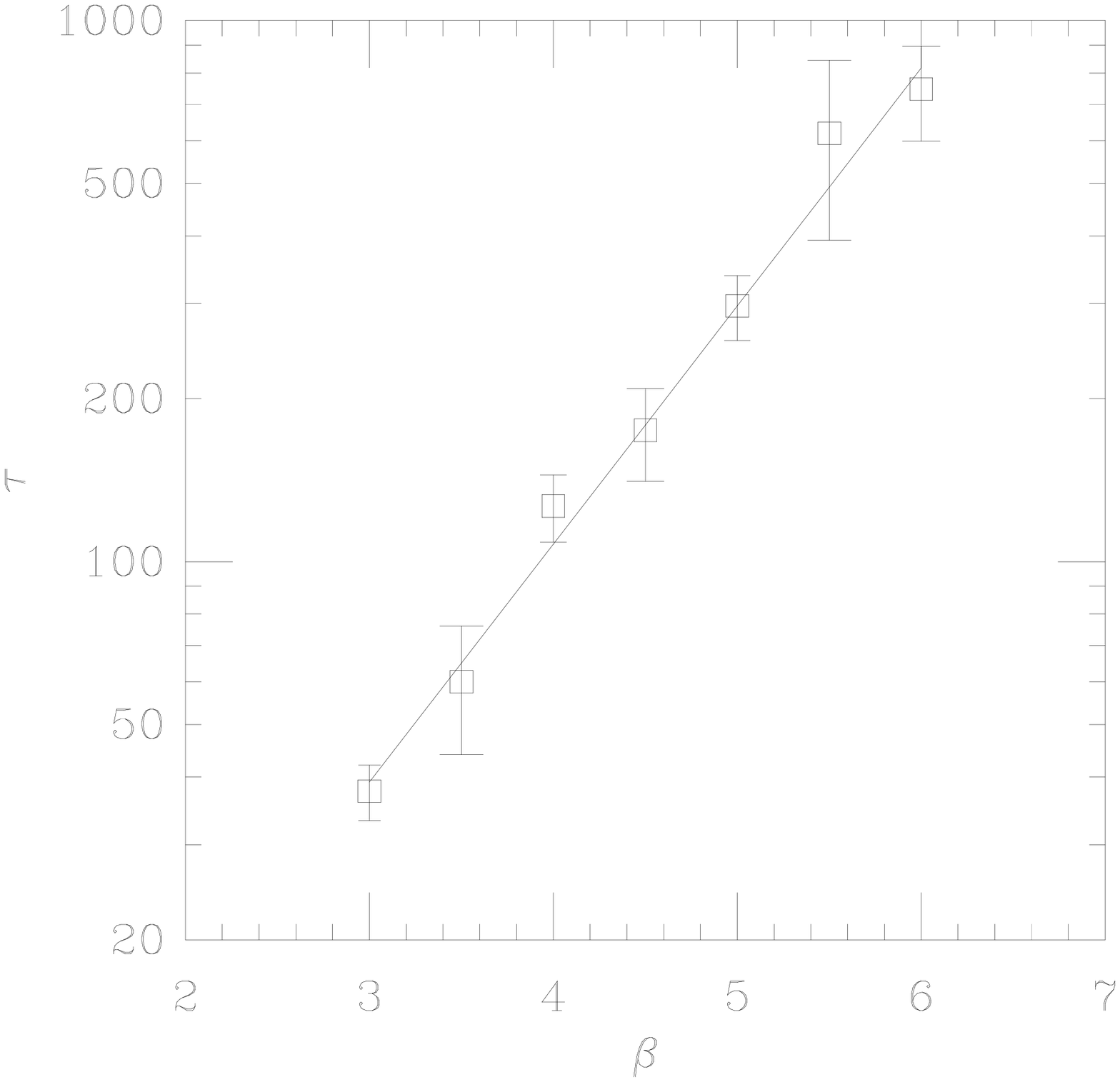}

\begin{references}
\bibitem[9]{Rt}In fact, the choice of phase for the Fourier transform
of $R(t)$ is arbitrary. There is a unique choice which would make
$R(t)$ causal. In the present work we choose zero phase for convenience.
\end{references}
\end{document}